\documentclass[11pt,letterpaper]{article}

\usepackage[margin=1in]{geometry}
\usepackage[T1]{fontenc}
\usepackage{lmodern}
\usepackage{microtype}
\usepackage{amsmath,amssymb,amsthm,mathtools}

\usepackage{graphicx}
\usepackage{booktabs}
\usepackage{enumitem}
\usepackage{caption}
\usepackage{subcaption}

\usepackage{algorithm}
\usepackage{algpseudocode}
\usepackage{placeins}

\usepackage[backend=biber,style=numeric-comp,sorting=nyt,maxbibnames=99]{biblatex}
\usepackage[hidelinks]{hyperref}
\usepackage[capitalise,nameinlink,noabbrev]{cleveref}

\newtheorem{theorem}{Theorem}[section]

\theoremstyle{definition}

\theoremstyle{remark}
\newtheorem{remark}{Remark}

\newcommand{\R}{\mathbb{R}}

\renewcommand{\P}{\mathbb{P}}

\title{Approximating the Girth of an Undirected, Weighted Graph for Larger Approximation Factors}

\author{
    Eliyahu Yablon\\
    Princeton University\\
    \texttt{yabloneli@gmail.com}
  }

\date{}

\begin{document}

\maketitle

\begin{abstract}
\noindent The girth of a weighted undirected graph is defined as the length of a shortest cycle in the graph. In this paper, we give a randomized $O\left(m^{1+2/(k-1)} n^{1/(k-1)}\right)$ algorithm to approximate the girth of a graph with $n$ vertices and $m$ edges to within a factor of $k$ for any real $k$. We assume that all edge weights are non-negative. For sufficiently high approximation factors, our algorithm outperforms the state of the art by polylogarithmic factors. Specifically, for approximation factors $k = c\frac{\log n}{\log \log n}$ where $c>5$, our algorithm gives a runtime of $O\left(m (\log n)^{5/c}\right)$, while instantiating the previous best $k$-approximation algorithms with $k = c\frac{\log n}{\log \log n}$ yields a running time of $O\left( \frac{cn (\log n)^{2 + 4/(3c)} }{\log \log n} + m\log n\right)$ \cite{kadria_et_al:LIPIcs.ESA.2026.82}. Furthermore, the algorithm presented here is conceptually simple, has relatively small constants, eschews single-source shortest path and sparse spanner algorithms, and employs a new technique, which we call ``random filtering.'' That being said, the basic algorithm returns only a numerical estimate for the girth, rather than a witness cycle. 
\end{abstract}

\section{Introduction}
\label{sec:introduction}
The girth of a graph is defined as the length of a shortest cycle. It is a fundamental graph parameter that has been studied algorithmically as far back as the 1970s.  A foundational algorithmic treatment of the problem was given by Itai
and Rodeh in 1978 \cite{ItaiRodeh1978} in which they gave an $O\left(\min(mn,n^{\omega})\right)$ algorithm to compute the girth of an unweighted, undirected graph, where $\omega$ is the matrix multiplication exponent. Since then, many advances have been made in computing graph girth. In 2016, Orlin and Sedeno-Noda gave an $O(mn)$ algorithm for computing the girth of a weighted, directed graph \cite{orlin2017nm}. However, the difficulty of exact girth computation and the inability to break the cubic runtime barrier for combinatorial algorithms have prompted researchers to instead explore approximation algorithms. 
\newline
\newline
\noindent The first truly subquadratic multiplicative approximation for
unweighted, undirected graphs was given by Lingas and Lundell, whose
randomized $\widetilde{O}(n^{3/2})$-time algorithm achieves an
$8/3$-approximation \cite{LingasLundell2009}. Roditty and Vassilevska
Williams later obtained the first subquadratic $2$-approximation: their
deterministic algorithm runs in $\widetilde{O}(n^{5/3})$ time
\cite{RodittyVassilevskaWilliams2012}. For directed graphs, Pachocki,
Roditty, Sidford, Tov, and Vassilevska Williams showed that, for every
integer $k\geq 1$, the girth of a directed graph with nonnegative integer
edge lengths can be approximated within a factor of $O(k\log n)$ in
\[
    O\!\left(mn^{1/k}\log^5 n\right)
\]
time with high probability \cite{pachocki2018approximating}. 
\newline
\newline
\noindent Most recently, Kadria, Roditty, and Vassilevska Williams obtained the
currently best general time--approximation trade-off for weighted
undirected graphs with nonnegative real edge lengths
\cite{kadria_et_al:LIPIcs.ESA.2026.82}. For every integer $k\geq 2$, their algorithm
computes a $(2k/3)$-approximation in
\[
    O\!\left(
        k n^{1+2/k}\log n
        +m(k+\log n)
    \right)
\]
time. This matches, up to polylogarithmic factors, the corresponding
trade-off previously known for unweighted graphs \cite{RodittyTrabelsi2025, KadriaRSWZ2022}.
\newline
\newline
\noindent Lastly, our algorithm is randomized and works by repeatedly drawing randomly sampled subgraphs of our graph. Our algorithm draws inspiration from other well-known randomized graph algorithms, such as Karger's algorithm \cite{karger1993global} and Alon-Yuster-Zwick's color-coding \cite{alon1995color}. Indeed, some of these randomized sampling techniques have even been used as subroutines in cycle algorithms before, such as in \cite{abboud2022hardness} where they make use of the observation that if each vertex is retained with probability $p$, then a given triangle is more likely to be in a subgraph than a 4-cycle (by precisely a factor of $p$).

\subsection{Our Contribution}
\label{subsec:contribution}
In this paper, we give a conceptually simple randomized algorithm that,
for any fixed constant failure probability $\delta$ and fixed $s\in \R^+$, gives a $\left(1+ \frac{\ln (8mn(2m+1))}{s} \right)$-approximation to the girth in time $O(e^s m \ln (1/\delta))$. The main advantages of our algorithm are
its simplicity, use of a new technique, and ability to approximate the girth to within any real-numbered factor. 
\newline
\newline
\noindent 
The key observation underlying our algorithm is that if we sample a random subgraph of $G$ by preserving each edge $e$ independently with probability $2^{-t\ell(e)}$, where $t$ is some parameter we can tune later, then the probability that a particular cycle appears in this subgraph decreases exponentially with its length. Thus, this sampling procedure strongly favors shorter cycles and can help us estimate the girth by changing the values of $t$ and seeing when cycles start to first appear with this sampling technique. The values of $t$ for which cycles start to appear are strongly controlled by the girth of the graph, allowing us to estimate the value of $g$. We refer to this technique as ``random filtering,'' and expect it to be useful for other problems in algorithmic graph theory (after all, this idea of sampling random subgraphs that favor particular ``small'' structures to appear can be applied to paths, trees, matchings, etc., and not just cycles). 

\begin{remark}[Changing the parameter $s$]
    If we just keep the parameter $s$ constant, we attain a strictly linear time algorithm for approximating the girth of a graph to within a logarithmic factor. By increasing the parameter $s$, we can improve our approximation factor at the cost of increasing our runtime, while decreasing the parameter $s$ degrades our approximation factor while improving runtime.
\end{remark}

\subsection{Notation}
Throughout this paper, any use of $\log$ or $\ln$ refers to the natural logarithm (though it usually doesn't matter since we deal with asymptotics). For any graph $G=(V,E)$, we let $V$ represent the set of vertices and $E$ represent the set of edges. To denote the weight of each edge, we consider the function $\ell : E \rightarrow \R _{>0}$ that assigns a positive weight to each edge. We assume without loss of generality that all edge weights are strictly positive; if any edge weights are zero, we first check to see if there is a zero-weight cycle. If not, for every zero-weight edge we could simply contract the two vertices connected by said edge without changing the girth of a graph (we preserve all loops and parallel edges in the contraction, so we are technically working with a multi-graph). Additionally, we will use the variable $g$ to denote the girth of a graph, and the function $d(u,v)$ to represent the minimum distance between two vertices, $u,v \in V$. Lastly, we assume that the graph we're working with has $2m \geq n$ and that there are no isolated vertices (we can just delete isolated vertices). For the bounds in terms of $n$, we assume $m\leq n^2$, as holds for simple input graphs; these input sizes are taken before any contractions that create multigraphs.

\section{The Algorithm}
\label{sec:algorithm}
\begin{algorithm}[h]
\caption{Random filtering with sampling parameter $s$}
\label{alg:approximate-girth}
\begin{algorithmic}[1]

\Require An undirected graph $G=(V,E)$ with positive edge lengths
$\ell$ and parameters $s>0$ and $\delta\in(0,1)$.
\Ensure With probability at least $1-\delta$, return $\widehat g$ where 
$g\leq\widehat g\leq(1+\ln(4A)/s)g$ and  $A=2mn(2m+1)$;
return $\infty$ if $G$ is acyclic.

\Function{ApproximateGirth}{$G,\ell,s,\delta$}
    \State $n\gets |V|$ and $m\gets |E|$

    \If{$G$ contains no cycle}
        \State \Return $\infty$
    \EndIf

    \State  $N\gets \lceil 12 e^s \ln(2/\delta) \rceil $
    \State $r\gets\left\lceil
    \frac{Ne^{-s}}{2} \right\rceil $
    \State $b \gets s + \ln (8mn(2m+1))$
    \State
    
    \For{$i=1,\ldots,N$}
        \State $T_i\gets\Call{SampleThreshold}{G,\ell}$
    \EndFor

    \State $\widehat T\gets$ the $r$-th largest of $T_1,\ldots,T_N$
        \Comment{Linear-time selection}
    \State $\widehat g\gets b/\widehat T$
    \State \Return $\widehat g$
\EndFunction

\end{algorithmic}
\end{algorithm}

\noindent The algorithm is described above. The main subroutine doing most of the work is \textsc{SampleThreshold}, though the description of what it's doing is very simple. Essentially, if for each edge, $e$, in our graph, we let $X_e \sim \text{Expon}(\ell(e))$, then we can define $H_t$ to be the subgraph of $G$ created by including all the edges $e$ of $G$ such that $X_e \geq t$. We can then define the random variable $T$, which is given by 
\[
T = \sup \{t \mid H_t \text{ has a cycle}\}
\]
All \textsc{SampleThreshold} does is return one number sampled from the probability distribution of $T$. Indeed, \textsc{SampleThreshold} can be computed in linear time, giving us a linear-time implementation of our algorithm. The exact method by which we do this is described in \cref{alg:sample-threshold}. Note that a naive implementation for \textsc{SampleThreshold} would sample the set $\{X_e \mid e\in E\}$ where $X_e\sim\operatorname{Exp}(\ell(e))$, and then sort the set $\{X_{e} \mid e\in E\}$ from smallest to largest. Afterwards, it would iterate through the set in descending order, adding the edges it is iterating over one by one to an initially edgeless graph until the first cycle appears. However, this takes $O(m\log n)$ time, since sorting a set of size $m$ takes at most this much time when $m\leq n^2$.
\newline
\newline 
\noindent To complete this task in linear time, perform the same sampling as before for each edge. Then find the median $M$ of the set $\{X_{e}\mid e\in E\}$, which can be done in linear time \cite{blum1973time}, and construct the graph $H_M$. If the number of edges is even, we take the upper median. If $H_M$ has a cycle, then $T\geq M$. Otherwise, $T<M$.
\newline
\newline 
\noindent 
If $H_M$ has a cycle, we may discard every edge $e$ such that $X_{e}<M$, since those edges cannot affect the largest threshold at which a cycle exists. We then recurse on the remaining edges.
\newline
\newline 
\noindent 
Now suppose that $H_M$ does not have a cycle, meaning that $H_M$ is a forest. Contract each connected component of $H_M$ into a single vertex, and delete every edge satisfying $X_{ab}\geq M$, since these are precisely the edges of $H_M$. Retain every edge satisfying $X_{ab}<M$, including edges whose endpoints are contracted to the same vertex. Such an edge becomes a self-loop and represents a cycle consisting of that edge together with the unique path between its endpoints in $H_M$. We must also retain parallel edges, since two parallel edges in the contracted multigraph may represent a cycle in the original graph. Thus, we can now recurse on the contracted graph, in which we now only have half as many edges, to find the value of $T$. 
\newline
\newline 
\noindent At each step, we recurse on a subgraph that has half as many edges, and each step takes linear time in the number of edges we currently have, so the total amount of time taken is
\[
m+m/2+m/4+\cdots=O(m).
\]

\begin{algorithm}[h]
\caption{Sampling the critical threshold $T$ in linear time}
\label{alg:sample-threshold}
\begin{algorithmic}[1]

\Function{SampleThreshold}{$G=(V,E),\ell$}
    \ForAll{$e\in E$}
        \State Sample $X_e\sim\operatorname{Exp}(\ell(e))$
    \EndFor
    \State \Return \Call{FindThreshold}{$G,X$}
\EndFunction

\Statex

\Function{FindThreshold}{$H=(W,F),X$}
    \If{$F=\varnothing$}
        \State \Return $-\infty$
    \ElsIf{$|F|=1$}
        \State \Return $X_e$ if $F=\{e\}$ and $e$ is a self-loop;
        otherwise return $-\infty$
    \EndIf

    \State $M\gets\operatorname{median}\{X_e:e\in F\}$
    \State $F^+\gets\{e\in F:X_e\geq M\}$ and
    $H^+\gets(W,F^+)$

    \If{$H^+$ contains a cycle}
        \State \Return \Call{FindThreshold}{$H^+,X|_{F^+}$}
    \Else
        \State Contract every component of $H^+$ in $H$, delete $F^+$,
        and retain all self-loops and parallel edges; call the result $H'$
        \State \Return \Call{FindThreshold}{$H',X|_{E(H')}$}
    \EndIf
\EndFunction

\end{algorithmic}
\end{algorithm}
\FloatBarrier

\section{Analysis}
\label{sec:Analysis}

\subsection{Bounding the Probability Distribution }
To make our observation in \cref{subsec:contribution} more rigorous and practical, we first prove \cref{thm:bounding}.
\begin{theorem}
\label{thm:bounding}
    For any undirected, weighted graph $G=(V,E)$, if for each edge $e\in E$ we sample a random variable $X_e \sim \text{Expon}(\ell(e))$, we let $H_t$ denote the subgraph of $G$ created by including all the edges $e$ of $G$ such that $X_e \geq t$, and we let $T = \sup \{t \mid H_t \text{ has a cycle}\}$, then 
    \[
    e^{-tg} \leq \P[T \geq t] \leq 2mn(2m+1) e^{-tg}
    \]
\end{theorem}
\noindent To prove these bounds, first note that the event $T\geq t$ is equivalent to the event that $H_t$ contains a cycle. The lower bound is immediate: the probability that $H_t$ contains a cycle is at least the probability that a minimum-weight cycle appears in $H_t$, which is $e^{-tg}$.
\newline
\newline 
\noindent 
The proof of the upper bound is adapted from proofs for established cycle counting bounds in graphs with high girth \cite{subramanian1995polynomial}, and also from cycle-freeness results for randomly sampled subgraphs with high girth \cite{putterman2026bounded}. First, note that between any two vertices in our graph, there is at most one simple path of length less than $g/2$. Indeed, if there were two distinct such paths, their union would contain a cycle of length less than $g$. Using this fact, for any graph we can non-constructively prove that there is a polynomial-sized collection of paths of length at least $g$ such that if $H_t$ were to contain a cycle, then it also must contain one of these paths. To show why such a family exists, note that for any cycle in $G$, we can start at some arbitrary vertex in the cycle, and traverse the edges along the cycle until the distance travelled is at least $g/2$. We can refer to this segment by $P_1e_1$, where $e_1$ represents the final edge along the segment and $P_1$ is the initial path along the segment. By construction, we must have that the length of $P_1$ is less than $g/2$. Now, if the length of $P_1e_1$ is greater than $g$, we are done. Otherwise, we can continue traversing along the cycle until the total accumulated length traversed is at least $g$. We can write this second segment as $P_2 e_2$ where the length of $P_2$ is less than $g/2$ and $e_2$ is some edge. Note that $P_1$ and $P_2$ are potentially empty paths. Thus, every cycle has an edge sequence of one of these forms 
\[
P_1 e_1 \qquad P_1e_1P_2e_2
\]
where the total weight of the edge sequence is at least $g$ and $P_1$ and $P_2$ both have length less than $g/2$. Now, note that there are at most $2mn$ edge sequences of the form $P_1 e_1$. This is because $P_1 e_1$ is perfectly specified by the starting vertex of the edge sequence, and the ending edge along with the orientation of the edge (recall that between any two points there is at most one path of length less than $g/2$ between them, so $P_1$ is entirely specified by its starting point and endpoint). There are $m$ edges, $2$ orientations per edge, and $n$ possible starting points, giving the upper bound of $2mn$. Using similar logic, we get that there are $4m^2n$ possible edge-sequences of the form $P_1e_1P_2e_2$, since there are $n$ starting points, $m$ possibilities for $e_1$, $m$ possibilities for $e_2$, and $2$ possible orientations per edge. This brings the total to $2mn + 4m^2n = 2mn(2m+1)$ possible edge sequences. Thus, if any cycle were to appear in $H_t$, we must have that at least one of these $2mn(2m+1)$ paths has appeared. The probability of any one of them appearing is at most $e^{-tg}$ since their lengths are at least $g$. Thus, 
\begin{align*}
    \P[T \geq t] &= \P [H_t \text{ has a cycle}] \\
    &\leq \P[\text{a path of the form } P_1e_1 \text{ or } P_1e_1P_2e_2 \text{ appears}]  \\
    &\leq 2mn(2m+1)e^{-tg}
\end{align*}
which completes the proof of \cref{thm:bounding}.

\subsection{Proving Correctness and Runtime}
For ease of notation, we will set $A=2mn(2m+1)$ in this section.
Now, to prove correctness of our algorithm, note that if we choose
any $s>0$ and set
\[
a=s \qquad b=s+\ln(4A),
\]
then, by the lower bound in \cref{thm:bounding},
\[
\P[T\geq a/g]\geq e^{-s},
\]
and by the upper bound in \cref{thm:bounding},
\[
\P[T\geq b/g]\leq \frac{e^{-s}}{4}.
\]
Thus, if we draw $N$ independent samples of $T$ and let
$\widehat{T}$ be the $\lceil Ne^{-s}/2\rceil$-th largest of these
samples, then
\begin{align*}
    \P\left[a/g\leq \widehat{T}\leq b/g\right]
        &\geq 1-\P[\widehat{T}<a/g]-\P[\widehat{T}>b/g] \\
        &\geq 1-\P\left[
            \operatorname{Binom}(N,e^{-s})< \lceil Ne^{-s}/2 \rceil 
        \right]
        -\P\left[
            \operatorname{Binom}(N,e^{-s}/4)\geq \lceil Ne^{-s}/2 \rceil 
        \right] \\
        &\geq 1-\exp\left(-\frac{Ne^{-s}}{8}\right)
                 -\exp\left(-\frac{Ne^{-s}}{12}\right) \\
        &\geq 1-2e^{-Ne^{-s}/12},
\end{align*}
where we used standard multiplicative Chernoff bounds in the
third line.
\newline
\newline
\noindent Thus, if we set $N=\lceil12e^s\ln(2/\delta)\rceil$,
we have that $a/g\leq\widehat{T}\leq b/g$ is satisfied with
probability at least $1-\delta$. Thus, we also have
\[
\frac{a}{\widehat{T}}\leq g\leq\frac{b}{\widehat{T}},
\]
meaning that if we output
\[
\boxed{\widehat{g}=\frac{s+\ln(4A)}{\widehat{T}}},
\]
then
\[
g\leq\widehat{g}\leq
\left(1+\frac{\ln(4A)}{s}\right)g,
\]
as desired.
\newline
\newline
\noindent We already noted that calling $\textsc{SampleThreshold}$
takes linear time. Computing the required order statistic takes
$O(N)$ time using linear-time selection, so the whole algorithm
takes $O(mN)=O(me^s\ln(2/\delta))$ time. In particular, the running
time is linear in $m$ for fixed $s$ and $\delta$.  
\newline 
\begin{remark}[A note on finite-precision implementations]
So far we have seemingly been assuming that our computer can manipulate real numbers and sample random variables perfectly. This assumption is not necessary for our algorithm to work; indeed, one can implement this algorithm in a word RAM model or even pointer-machine model where the only sacrifice is a small increase in the error probability. For exact details on how this can be done one can simply ask their favorite frontier large language model. However, if we just implement the exponential filtering algorithm using 64-bit float arithmetic, then some back-of-the-envelope calculations reveal that the increase in error probability is negligible. To see this, if we let $u=2^{-53}$ (a float has 53 bits of significand precision), then we can generate exponential random variables with rate $\ell (e)$ by sampling a uniform value, $U$, in $\{u, 2u, \ldots, 1-u\}$ and computing $\frac{-\ln U}{\ell (e)}$. Assuming our logarithm and division operations are correct to within a relative error $u$, then the rounding errors from this operation only introduce an error of $O(2^{-53})$ in the CDF of our random variable, and the errors from the discretization of our probability sampling introduce an error of at most $2^{-53}$ in the CDF, meaning we only have an error of $O(2^{-53})$ in our CDF per sampled random variable. The rest of the rounding errors in our calculations will also incur an $O(2^{-53})$ error, which is all negligible assuming reasonably sized graphs (i.e., not on the order of $2^{53}$) and edge weights (i.e., no risk of overflow or underflow).  
\end{remark}

\subsection{Tuning the Parameter $s$}
In this subsection we'll explain how one can tune the parameter $s$ to get a desired approximation ratio. Suppose one wants to approximate the girth of a graph to within a factor $k$. This means that 
\[
1+ \frac{\ln (8mn(2m+1))}{s}  = k \implies s = \frac{\ln (8mn(2m+1))}{k-1}
\]
Substituting this into our runtime formula yields 
\begin{align*}
    e^sm &= e^{\frac{\ln (8mn(2m+1))}{k-1}} m  \\
    &\leq 16^{\tfrac{1}{k-1}} ((m+1)mn)^{\tfrac{1}{k-1}}m \\
    &= O\left(m^{1+\tfrac{2}{k-1}} n^{\tfrac{1}{k-1}}\right)
\end{align*}
Thus, if we then set $k = c\frac{\log n}{\log \log n}$ where $c>5$ is fixed, we get that 
\begin{align*}
m^{1+\frac{2}{k-1}}n^{\frac{1}{k-1}}
&\le m\exp\left(\frac{5\log n}
{c\frac{\log n}{\log\log n}-1}\right)\\
&=m\exp\left(
\frac{5\log n\log\log n}
{c\log n-\log\log n}
\right)\\
&=m\exp\left(
\frac{5\log\log n}
{c-\frac{\log\log n}{\log n}}
\right)\\
&=m(\log n)^{
\frac{5}{c-\frac{\log\log n}{\log n}}
}\\
&=m(\log n)^{5/c+O(\log\log n/\log n)}\\
&=O\left(m(\log n)^{5/c}\right)
\end{align*}
where on the first line we used $m\leq n^2$. The final step uses $\exp\!\left(O((\log\log n)^2/\log n)\right)=O(1)$. These guarantees match the ones given in the abstract.

\section{Conclusion and Future Directions}
We presented an algorithm that can approximate the girth of an undirected, weighted graph to within any factor $k>1$, though it only improves upon the state of the art algorithms for $k > 5\frac{\log n}{\log \log n}$. Additionally, the algorithm is simple and very easy to implement, while also utilizing a technique not currently used in state-of-the-art algorithms. For future papers, it'd be interesting to apply this random filtering framework to other algorithmic graph theory problems. Namely, it'd be interesting to see if there are other problems in which one can design a random sampling method that heavily favors ``small'' constructions to appear and then leverage this to compute desired quantities of the graph. It'd also be interesting to explore if the girth approximation algorithm presented in this paper lends itself to more efficient parallelized algorithms. The main bottleneck in this algorithm's runtime is that one needs to draw many random samples from the same distribution, but this could easily be parallelized. Lastly, the main limitation of this algorithm is that it doesn't return a witness cycle. If we let $T$ be defined as in \cref{thm:bounding}, then perhaps we can recover a witness cycle by looking at the weight of the only cycle present in $H_T$ and recording the smallest weight across all our trials. After all, the weight of the cycle in $H_T$ intuitively seems to be a crucial piece of information about our graph, yet it is completely neglected in our algorithm. 

\newpage 
\subsection*{Acknowledgments}
Many thanks to Professor Vassilevska Williams and Professor Tarjan for helpful questions and discussions related to this research. 
\newline
\noindent 
\textit{AI Acknowledgement.}
ChatGPT5.6sol (high thinking) was used throughout the development and writing of this paper. The authors developed the paper's core insight and intuition: sampling random subgraphs so that longer cycles are exponentially less likely to appear, and then extracting information from the point at which cycles first begin to occur. ChatGPT5.6sol was used for much of the technical work that followed, including the linear-time implementation of \cref{alg:sample-threshold}, the formulation and proof of the upper bound in \cref{thm:bounding}, and the replacement of Bernoulli random variables with exponentially decreasing success probabilities by an equivalent formulation using exponential random variables. The approximation ratio and upper bound in theorem 3.1 was then optimized further by ChatGPT6Astra (Max thinking).

\printbibliography

\end{document}